# Crystal-lattice coupling to the vortex-melting transition in YBa$_2$Cu$_3$O$_{7-\delta}$

R. Lortz, [1,2] C. Meingast, [1] U. Welp, [3] W. K. Kwok, [3] G. W. Crabtree [3]

[1] *Forschungszentrum Karlsruhe, Institut für Festkörperphysik, 76021 Karlsruhe, Germany*
[2] *Fakultät für Physik, Universität Karlsruhe, 76131 Karlsruhe, Germany*
[3] *Materials Science Division and Science and Technology Center for Superconductivity, Argonne National Laboratory, Argonne, Illinois, U.S.A*

Distinct discontinuities in the thermal expansion of the crystal lattice are observed at the melting-transition of the vortex lattice in a naturally untwinned reversible YBa$_2$Cu$_3$O$_{7-\delta}$-single crystal using high-resolution dilatometry. This coupling between the vortex transition and the crystal lattice demonstrates that the crystal lattice is more than a mere host for the vortices, and it is attributed to a strong pressure dependence of the superconducting transition temperature and thus to the condensation energy at the vortex melting temperature.

High-T$_c$ superconductors are a unique class of materials in which the vortex system can display a wide variety of different phases [1-3]. These phases and the transitions between them can be controlled by temperature, applied magnetic field and pinning centers [1-3]. Of particular interest is the vortex-lattice melting transition observed in clean, low-pinning YBa$_2$Cu$_3$O$_{7-\delta}$ and Bi$_2$Sr$_2$CaCu$_2$O$_{8+\delta}$ crystals [3-13]. This is a first order transition leading from the Abrikosov vortex lattice state to the vortex liquid state in a fashion very analogous to the melting of ordinary matter. In fact, by measuring the latent heat and the discontinuity in the magnetization, that is, in the vortex spacing, thermodynamic consistency was demonstrated when using the Clausius-Clapeyron equation [4,6]. In this scenario the crystal lattice serves merely as a host for the interacting vortices, and models based on the Lindemann criterion [14] in which the lattice will melt if the thermal displacements of the particles (vortices) reach a certain fraction of the lattice constant have been successful in explaining the observed results [2,15,16]. This could be expected since in clean, low-pinning YBa$_2$Cu$_3$O$_{7-\delta}$ and Bi$_2$Sr$_2$CaCu$_2$O$_{8+\delta}$ crystals the dominant coupling between vortices and the crystal lattice, namely pinning [17], is absent.

In this Letter we show that distinct discontinuities in the thermal expansion of the *crystal lattice* occur at the vortex lattice melting transition of YBa$_2$Cu$_3$O$_{7-\delta}$, which demonstrates that the crystal lattice is more than just a host for the vortices. These discontinuities are positive (negative) along the b-axis (a-axis) and track closely those observed at the zero-field T$_c$ [18-20]. We attribute this coupling between the vortex transition and the crystal lattice to a strong pressure dependence of the underlying electronic structure responsible for superconductivity. This shows that unlike in the melting transition of ordinary matter, where the electronic structure usually only plays a minor role and the transition is largely driven by the configurational entropy, the superconducting condensation energy of the underlying crystal host cannot be neglected in the vortex system, in agreement with recent theoretical calculations of vortex melting [21-23].

The same untwinned YBa$_2$Cu$_3$O$_{7-\delta}$ single-crystal with dimensions L$_a$ x L$_b$ x L$_c$ =1.06 x 0.83 x 0.64 mm$^3$, which was used in previous specific heat measurements by Schilling et al. [4], was used for the present study. This crystal shows a pronounced specific heat peak at the vortex-lattice melting, as well as fully reversible behavior in a large temperature interval around T$_m$ [24]. The thermal expansion measurements were performed with a capacitance dilatometer in a continuous-heating mode with a rate of 18 mK/s. Data points were taken every 0.03 K. Due to the small size of the crystal and the small magnitude of the anomalies at the melting transition, the data of 6-10 heating cycles were averaged and then further smoothed over 10 points in order to improve the signal-to-noise ratio.

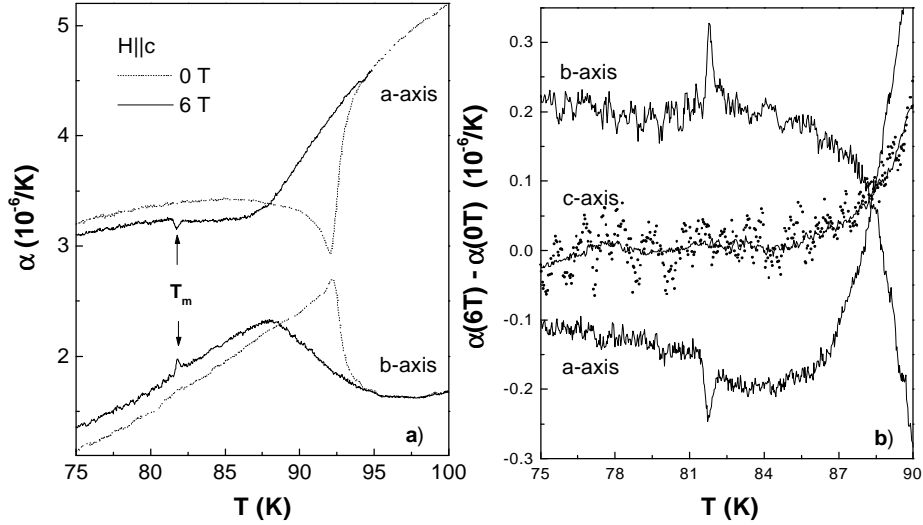

FIG. 1. a) Thermal expansivity of the a- and b-axis in H=0 T and in H=6 T applied parallel to the c-axis and b) for the a-, b- and c-axis in H=6 T after subtraction of the zero-field data as a background. Clear peak-like features are seen at the vortex-melting transition at $T_m$=81.7 K in the 6 T data for the a- and b-axes.

FIG. 1a shows the resulting expansivity data for the orthorhombic a- and b-axes for fields of 0 T and 6 T applied along the c-axis. 6 T was chosen because the specific heat shows the largest entropy jump at this field [4]. The zero-field data exhibit sharp lambda-type transitions at $T_c$=92.4 K of opposite sign in the a- and b-axes, which have previously been analyzed in detail [18-20,25]. In a field of 6 T the expansivity anomalies at $T_c$ are broadened very much like the specific heat anomaly [4,10,26], and, in addition to the zero-field anomaly, small sharp peaks are seen in both axes at $T_m$=81.7 K, which is precisely the temperature where melting is seen in the specific-heat measurements [4]. The details of the transitions are more clearly seen in Fig. 1b, where the zero-field data have been subtracted from the 6 T data. The melting peaks in the thermal expansivity are of opposite sign in the a- and b-axis and appear to correlate with the sign of the anomalies at $T_c$. No anomaly (within the limit of our resolution) could be detected in the c-axis data, in which the anomaly at $T_c$ is also much smaller. The best data are for the a-axis, and the shape of this anomaly is very similar to the specific heat anomaly, i.e. there is a peak and a jump in the expansivity. The peak in the thermal expansion is somewhat broader than in the specific heat due to the averaging procedure. We note that vortices can couple to the crystal lattice through pinning forces, which can result in quite large irreversible magnetostrictive effects [17]; this can however be ruled out here because of the reversible nature of the present crystal, which allows the following thermodynamic treatment of our data.

The transition of the superconductor at $T_m$ will in general depend on pressure, P, and on magnetic field, H, and the following Clausius-Clapeyron equations

$$dT_m / dP_i = (\Delta L_i / L_i) \cdot V_{mol} / \Delta S \qquad - 1 -$$

$$dT_m / dH = (\Delta B) \cdot V_{mol} / \Delta S \qquad - 2 -$$

are expected to hold since the transition is first-order. In Eq. 1, the index 'i' refers to the index of the a-, b- or c-axis of the orthorhombic system, and $\Delta L_i$ and $\Delta S$ are the length and entropy jumps at the transition. The thermodynamic consistency of Eq. 2 has been demonstrated with specific heat and magnetization data [4,6]. Eq. 1 can be used to calculate the uniaxial $dT_m/dp_i$ values using our measured length jumps ($\Delta L_a/L_a$= (2.5±0.25)·$10^{-8}$ and $\Delta L_b/L_b$ = (2.1±0.5)·$10^{-8}$) and the entropy jump ($\Delta S$= (1.25±0.2)·$10^{-3}$ J/(mol·K)) from Ref. [24]) at H=6 T, and the results are listed in

Table I together with the uniaxial pressure dependencies of $T_c$ calculated via the Ehrenfest equation ($dT_c/dp_i = \Delta\alpha_i \cdot T_c \cdot V_{mol}/\Delta C_p$).

We find that the values of $dT_m/dp_i$ and $dT_c/dp_i$ are approximately equal. (We note that the presently derived $dT_c/dp_i$ values exhibit the same a,b anisotropy as those previously determined via thermal expansion [20,25] and direct pressure [27].)

| $i=$ | $dT_m/dp_i$ | $dT_c/dp_i$ |
|---|---|---|
| $a$ | - 2.1 ± 0.5 K/GPa | - 2.7 ± 0.02 K/GPa |
| $b$ | + 1.8 ± 0.7 K/GPa | + 1.9 ± 0.08 K/GPa |
| $c$ | 0 ± 1.0 K/GPa | - 0.9 ± 0.6 K/GPa |

TABLE I. Uniaxial pressure coefficients for $T_m$ and $T_c$ derived from the present thermal expansivity data and specific heat data from Ref. [4].

The crystal-lattice response at $T_m$ in a sense occurs because $T_m$ is pressure dependent (Eq. 1), and it is interesting to ask the question: what is the mechanism for this pressure dependence? Pressure is not expected to couple directly to the vortex lattice, e.g. by changing the vortex-vortex distance (which can be tuned nicely with the magnetic field). Pressure does, however, change the crystal lattice, which in general will affect fundamental superconducting parameters such as $T_c$, $H_c$, $\kappa$ or $\lambda$. Turning this idea around - the length changes of the crystal lattice at the melting transition can thus be viewed to result from changes in the superconducting parameters at $T_m$. This is supported by calculations of the entropy jump at $T_m$, which suggest that most, if not all, of the entropy result from changes in the superconducting parameters rather than from configurational entropy of the vortices [21-23].

The close correspondence between $dT_m/dp_i$ and $dT_c/dp_i$ (Table 1) suggests that the pressure dependence of $T_c$ plays the crucial role for $dT_m/dp_i$, and a quite natural way to obtain this relationship is through a power law of the sort

$$H_m = a\left(1 - T_m/T_c\right)^{2n} \qquad - 3 -$$

which directly links $T_m$ to $T_c$. From this it follows naturally that if $T_c$ is raised (lowered) by applying uniaxial pressure, the melting line will be shifted to higher (lower) temperatures by roughly the same amount. Eq. 3 has been shown to describe the field-dependence of $T_m$ quite well using e.g. a 3d-XY ($\nu=0.669$) exponent [10,11,28]. The effect of applying pressure may be formally characterized by pressure dependent values of $T_c$ and/or $a$. The value of $\nu$ should remain unaffected, since it is 'universal' in a scaling approach. We note, that in the 3d-XY approach Eq. 3 results from the fact that the vortex lattice parameter $a_0$ is proportional to $H^{-1/2}$ and the coherence length follows a power law: $\xi=\xi_0|t|^{-\nu}$ ($t=(1-T/T_c)$); vortex melting occurs when the coherence length reaches some fixed fraction of the vortex-vortex distance. Differentiating $T_m$ in Eq. 3 with respect to uniaxial pressure yields:

$$\frac{dT_m}{dp_i} = \left[1 - \left(\frac{H_m}{a}\right)^{\frac{1}{2n}}\right]\frac{dT_c}{dp_i} + \left[\frac{T_c H_m^{\frac{1}{2n}}}{2n}a^{-\left(\frac{1}{2n}+1\right)}\right]\frac{da}{dp_i} \qquad - 4a -$$

For the present case ($H_m=6$ T, $T_m=81.7$ K, $T_c=92.3$ K, $a=108.6$ T), Eq. 4a reduces to:

$$\frac{dT_m}{dp_i} \approx 0.9\frac{dT_c}{dp_i} + \left(7.3 \cdot 10^{-2}\frac{K}{T}\right)\frac{da}{dp_i} \qquad - 4b -$$

The strong correlation between $dT_m/dp_i$ and $dT_c/dp_i$ (Table 1) suggests that the second term in Eq. 4b is small, and, therefore, that the primary factor determining $dT_m/dp_i$ are the zero-field $dT_c/dp_i$ values.

The parameter $a$ in $YBa_2Cu_3O_{7-\delta}$ on the other hand, determines the field scale of the melting line, and, for H applied parallel to the c-axis, is determined entirely by the anisotropy of the superconductor [10,12,29,30]. In the following we make some simple calculations of the pressure dependence of $a$, which indeed show that this effect can be neglected for $YBa_2Cu_3O_{7-\delta}$.

Anisotropy depends strongly on the crystal structure [31] and on the doping level [29], both of which may be affected by pressure [32]. For example, by changing the oxygen content of a $YBa_2Cu_3O_{7-\delta}$ sample from $\delta=0.0$ to $\delta=0.06$ one decreases $a$ by ~24% [10,12], which is correlated to a change in anisotropy from 5.3 to 7.0 and a change in hole concentration of $\Delta n_h \approx 0.02$ [33]. Assuming that pressure changes the anisotropy and hole concentration in the same fashion as oxygen doping, we can calculate the effect of pressure on the anisotropy using the uniaxial charge-transfer coefficients $dn_h/dp_a=0.0024$ GPa$^{-1}$, $dn_h/dp_b=$

-0.0008 GPa$^{-1}$ and $dn_h/dp_c$=0.0017 GPa$^{-1}$, which were determined using thermal expansion data within a simple pressure-induced charge-transfer model [32,34] using Eq. 5,

$$\frac{dT_m(n_h)}{dp} = \left(\frac{dT_m}{dn_h}\right)_p \cdot \left(\frac{dn_h}{dp}\right)_T \qquad - 5 -$$

We find $dT_m/dp_a$=-0.24 K/GPa, $dT_m/dp_b$= +0.08 K/GPa and $dT_m/dp_c$= -0.17 K/GPa, which are all significantly smaller than the corresponding $dT_c/dp_i$ values, so that anisotropy changes through pressure-induced charge-transfer should only play a minor role in determining $dT_m/dp_i$. The anisotropy change due to pressure-induced changes in the crystal structure can be estimated using the results of Tallon et al. [31], in which he showed that the irreversibility field (or melting line) follows a very simple exponential dependence $H_m=H_{m0}*\exp(-d_b/\xi)$, where $d_b$ is the blocking layer distance and $\xi$ is a coherence length. The increase of the melting field due to this effect should be largest for c-axis pressure and can be calculated using

$$\frac{da}{dp_c} = \frac{dH_m}{dp_c} = \left(\frac{dH_m}{dc}\right)_p \left(\frac{dc}{dp_c}\right)_T \qquad - 6 -$$

and the c-axis compliance $1/c*dc/dp$= 4.7 x 10$^{-3}$/GPa [35]. We find $dT_m/dp_c$= 1.8 x 10$^{-2}$ K/GPa. This calculation, in which it was assumed that pressure-induced changes in $d_b$ scales with the total c-axis, shows that this effect is at least an order of magnitude smaller than the $dT_c/dp_i$ values. Thus, structural induced changes in the anisotropy are not expected to play a significant role in the uniaxial pressure effects of $T_m$ in $YBa_2Cu_3O_{7-\delta}$. We note that in contrast to $YBa_2Cu_3O_{7-\delta}$, the hydrostatic pressure dependence of the irreversibility line ($dT_{irr}/dp_{hydr}$=2 K/GPa) of $Bi_2Sr_2CaCu_2O_{8+\delta}$, which has a much 'softer' c-axis, can be attributed to the pressure induced change in anisotropy [36].

Summarizing, a clear expansion (contraction) of the crystallographic b-axis (a-axis) is observed at the vortex-melting transition in $YBa_2Cu_3O_{7-\delta}$ using high-resolution dilatometry. This response of the crystal lattice, which is in accord with the thermodynamical expectations based on the Clausius-Clapeyron equation and on the assumption that $T_m$ follows a power law of the form as in Eq. 3, shows that the crystal is more than just a host for the vortices. Physically, this response can be traced back to the large uniaxial pressure dependencies of $T_c$ and, thus, of the condensation energy at $T_m$. Our data provide a third (besides magnetization and specific heat) independent thermodynamic consistency check of the melting transition in the sense that the entropy jump from the specific heat data provides a very reasonable value of $dT_m/dp_i$ using the Clausius-Clapeyron equation.